\documentclass{Interspeech2024}

\interspeechcameraready

\title{Interface Design for Self-Supervised Speech Models}

\name[]{Yi-Jen}{Shih}
\name[]{David}{Harwath}

\address{
  Department of Computer Science, The University of Texas at Austin, USA
  }
\email{yjshih@utexas.edu, harwath@utexas.edu}


\usepackage{multirow,makecell}

\usepackage{xcolor}

\usepackage{soul}

\begin{document}

\maketitle

\begin{abstract}
    Self-supervised speech (SSL) models have recently become widely adopted for many  downstream speech processing tasks. The general usage pattern is to employ SSL models as feature extractors, and then train a downstream prediction head to solve a specific task. However, different layers of SSL models have been shown to capture different types of information, and the methods of combining them are not well studied. To this end, we extend the general framework for SSL model utilization by proposing the interface that connects the upstream and downstream. Under this view, the dominant technique of combining features via a layerwise weighted sum can be regarded as a specific interface. We propose several alternative interface designs and demonstrate that the weighted sum interface is suboptimal for many tasks. In particular, we show that a convolutional interface whose depth scales logarithmically with the depth of the upstream model consistently outperforms many other interface designs.
\end{abstract}

\section{Introduction}
\label{sec:intro}
Speech processing tasks such as automatic speech recognition 
have traditionally relied on supervised learning algorithms that require transcribed training data.
However, because transcriptions are expensive and time-consuming to collect, self-supervised speech models have recently become extremely popular. 
Speech models based on self-supervised learning (SSL) can be pre-trained with large amounts of unlabeled data, e.g. using a masked language modeling objective, and subsequently fine-tuned on a relatively small amount of labeled data for a target downstream task. 
Recent literature has proposed various SSL algorithms~\cite{hsu2021hubert,Chen2021WavLMLS,Conneau2020xlsr,schneider2019wav2vec} and many applications have been built on top of these SSL speech models~\cite{mohamed2022self}, including Automatic Speech Recognition~\cite{baevski2021wav2vec-u,liu2022wav2vec-u2}, Visually Grounded Speech~\cite{peng2022fastvgs,peng2022vghubert,shihSpeechCLIP,layne_mspchclip}, Speech Segmentation~\cite{peng2022vghubert,Strgar_2022}
Emotion recognition~\cite{Morais_emotion_ssl}, speaker verification~\cite{Peng_SSL_sv_2023} and so on.
Furthermore, there are benchmarks proposed to evaluate speech SSL models on various downstream tasks~\cite{yang21c_superb,tsai2022superbsg,shon-etal-2023-slue,shi23g_mlsuperb}.

\begin{figure}[t]
    \centering
    \includegraphics[width=\linewidth,trim={3.7cm 0.25cm 2.1cm 0},clip]{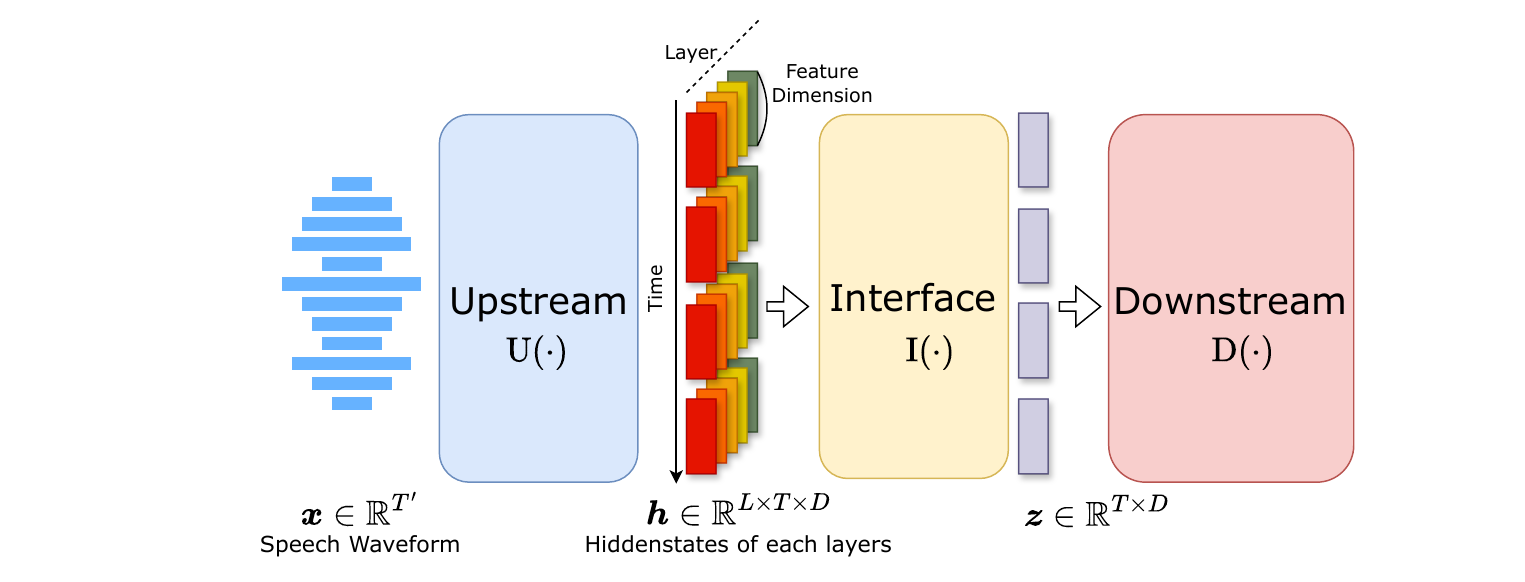}
    \caption{The general framework for utilizing self-supervised speech models only considers the upstream and downstream model components. We argue that the interface connecting them should be considered in its own right as a separate component.($L$~:~number of upstream model layers,~$T$~:~upstream model output sequence length,~$D$~:~upstream model feature dimension)}
    \label{fig:framework}
\end{figure}

While significant progress has been made in developing algorithms and applications of speech SSL models, it is still unclear what the best method is for utilizing these models on downstream tasks.
Existing methods for utilizing speech SSL models fall roughly into 3 categories:
1) Cascade the upstream model with a task-specific downstream prediction head and fine-tune the entire model.
2) Freeze the upstream model and take the output of a specific layer as the input features for a task-specific downstream module, then train the downstream module.
3) Compute a learnable weighted sum of all layers belonging to the upstream model and use the result as the input features for a task-specific downstream module. Typically, only the summation weights and the downstream module are trained during fine-tuning.
Among the three methods, Fine-tuning the upstream and downstream models together often leads to better performance, but it is far more computationally expensive and can be unstable when a small amount of fine-tuning data is available.
In contrast, using features from a single layer of the frozen upstream model is the most computationally inexpensive, however, it neglects the non-uniform distribution of different information types encoded across different layers of the upstream model~\cite{Ankita2022CCA}.
The weighted sum can thus be viewed as an intermediate trade-off between the other two options and is the default configuration used in the SUPERB~\cite{yang21c_superb} benchmark.

However, in this paper, we argue that the weighted sum is still a suboptimal way of combining information across layers of speech SSL models.
Our intuition is that because the information encoded in each dimension might differ across each layer of the upstream model, naively summing over the layers dimension-by-dimension may lead to an information loss in proportion to the degree of statistical independence between the same dimension across multiple layers.

Built upon this motivation, we hypothesize that there exists a better way of aggregating the information across different layers and introducing an \textbf{Interface} module in the SSL framework.
The overall framework for utilizing speech SSL models then becomes a set of three components: the
\textbf{Upstream} model, the \textbf{Downstream} prediction head, and the \textbf{Interface} that bridges between them by aggregating information across all upstream model layers.
Under this definition, the widely-used weighted sum method is a specific type of interface, and in this paper, we further propose several alternative interface designs.
We evaluate the performance of each proposed interface across 5 upstream speech SSL models and several downstream tasks in the ML-SUPERB~\cite{shi23g_mlsuperb} and SUPERB~\cite{yang21c_superb} benchmarks.
Among all our proposed interfaces, we show that Hierarchical Convolution across layers tends to achieve the best overall performance.
We also conduct additional experiments showing that performance differences among various interface designs are due to their architectures themselves, and are not merely due to the difference in the number of trainable parameters.
Additionally, we show that the Hierarchical Convolution interface still outperforms the weighted sum interface when the upstream model is fine-tuned with the downstream model.
The code is publicly available.
\footnote{\url{https://github.com/atosystem/SSL_Interface}}
Our contributions in this paper can be summarized as follows:
\begin{enumerate}
    \item We propose an updated framework for self-supervised speech models that formally recognizes the interface module
    \item We introduce several designs for the interface and show that the Hierarchical Convolution tends to perform the best across many speech processing tasks
    \item We show that even when the full model is end-to-end fine-tuned, different interfaces still show performance differences.
\end{enumerate}

\section{Proposed Interface Methods}
\label{sec:method}
\begin{figure*}[t]
    \centering
    \includegraphics[width=\linewidth,trim={0cm 0.20cm 0cm 0},clip]{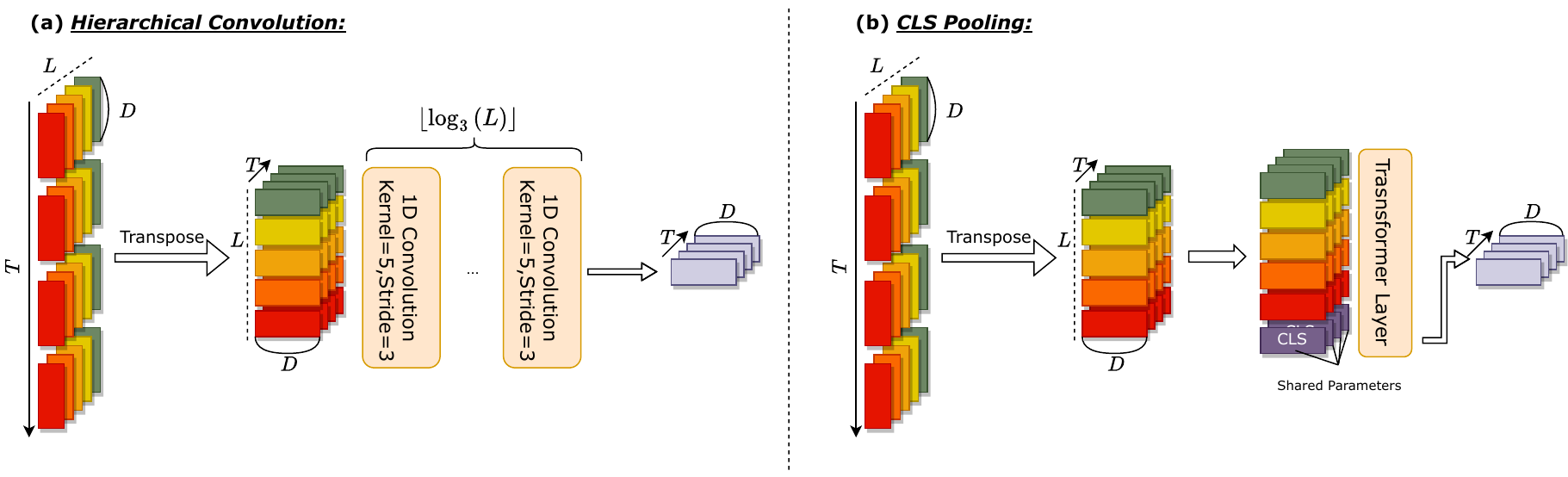}
    \caption{Proposed Interface designs. (a): Hierarchical Convolution is applied over the layer dimension of the upstream model hiddenstates. (b): A learnable CLS embedding is used for summarizing over layer dimension}
    \label{fig:interface_conv_cls}
    \vspace{-10pt}
\end{figure*}

\subsection{Interface Definition}

We formalize our proposed overall pipeline for utilizing self-supervised speech models as a general framework (as shown in Fig.~\ref{fig:framework}): \textbf{Upstream} $\rightarrow$ \textbf{Interface} $\rightarrow$ \textbf{Downstream}.
The \textbf{Upstream} model $\mathrm{U}\left(\cdot \right): \mathbb{R}^{T'} \mapsto \mathbb{R}^{L\times T\times D} $ stands for a pre-trained, self-supervised speech model which maps an utterance waveform ${x} \in \mathbb{R}^{T'}$ of length $T'$ to a hidden representation $\boldsymbol{h}$ which consists of $L$ representation layers each with a feature dimension of $D$.
\textbf{Interface} is a function that aggregates information over the layer dimension $L$ of the upstream output, $\mathrm{I}\left(\cdot \right): \mathbb{R}^{L\times T\times D} \mapsto \mathbb{R}^{ T\times D} $.
Finally, \textbf{Downstream} $\mathrm{D}\left(\cdot \right)$ takes $\boldsymbol{z} = \mathrm{I}\left( \boldsymbol{h} \right)$ as input and outputs predictions for a specific downstream task.
Under this general framework, the weighted sum can be viewed as a specific type of interface,
\begin{equation}
    \mathrm{I}_{\mathrm{WS}}\left( \boldsymbol{h}  \right)=\sum_{l}^{L}{w_l \cdot \boldsymbol{h}_l}
    \label{eq:weightedsum}
\end{equation}
where $\boldsymbol{w}$ is the per-layer weight vector, which is typically learned jointly with the downstream model weights.

The weighted sum interface has the benefit of being able to aggregate information across all $L$ upstream layers with only $L$ learnable parameters required.
On the other hand, we hypothesize that it has potential downsides: by directly summing layer representations dimension-wise, statistically independent features from different layers of the upstream model may "collide" with one another resulting in information loss.
As the number of layers in the upstream model increases, this problem can become more severe.
To further investigate this, we propose several alternative interface designs that attempt to avoid information collision across multiple upstream layers while remaining relatively lightweight and also keeping the output dimension roughly the same as the upstream model's dimension.


\subsection{Proposed Interfaces}

\subsubsection{Grouped Weighted Sums}
An intuitive way to reduce the information collision of a single weighted sum is instead to use multiple weighted sums, each of which only has access to a subset of the upstream layers.
Specifically, after conducting a weighted sum within each subset of upstream layers, we concat the results of each group along the feature dimension.
Then we use a learnable projection to project the feature dimension to that of the downstream model.


\subsubsection{Concatenation + Learnable Projection}

Another type of interface concatenates all layer representations $\boldsymbol{h}$ along the feature dimension and uses a learnable projection layer to automatically decide the features that are beneficial to the downstream model. 
Specifically, we reshape $\boldsymbol{h}$ from $\left( L,T,D \right)$ to $\left( T,L\times D \right)$,
then 
project $\left( T,L\times D \right)$ back to the original upstream model dimension $\left( T, D \right)$.



\subsubsection{Hierarchical Convolution over Layers}

Even though the feature manifold differs among each layer of the upstream model, neighboring layers can still possess similar feature distributions due to the residual connection in the transformer layers~\cite{vaswani2017attention}.
Motivated by the local structure of the hidden representation across the layer dimension, we apply 1D convolutions over the layer dimension to aggregate the information across layers.
In practice, an interface design should be adaptable to upstream models with an arbitrary number of layers.
Hence, in our paper, we fix the kernel size=$5$, stride=$3$ for our convolution layer and we stack $\lfloor \mathrm{log}_3{L} \rfloor$ identical convolutional layers, which has the effect of collapsing the output of all models layers down to a single vector at the output of the convolutional stack (as shown in Fig.~\ref{fig:interface_conv_cls} (a)).
We leave further convolutional hyperparameter optimization for future study.

\subsubsection{CLS Pooling over layer dimension}

We can use an attention module to aggregate information across layers rather than using convolutions.
Specifically, at each timestep we view all the layers of the upstream model as a sequence of tokens and concatenate a learnable CLS vector~\cite{devlin-etal-2019-bert} to this sequence. A Transformer layer is then used to aggregate information across all layers at the same time step into the CLS token (as shown in Fig.~\ref{fig:interface_conv_cls} (b)). 
A key difference compared to our other proposed interface designs is that CLS Pooling is data-dependent, while other methods are only task-dependent. 

\subsubsection{PCA + Concatenation}

Besides the above parametric methods, we also investigated layer-wise dimensionality reduction via Principal Component Analysis~(PCA), followed by concatenation across layers. Using the HuBERT Base model as an example upstream model, we take only the top 60 principal components of each Transformer layer and concatenate them together, resulting in $60\times 13=780 \approx768$ dimensions, which is roughly the same as the upstream model feature dimension. 
Practically, the PCA transform is learned from each downstream task dataset respectively.

\section{Experiments}
\label{sec:evaluation}
\begin{table}[th]
  \caption{Comparison between different proposed interfaces with HuBERT Base as the upstream model, evaluated on the ML-SUPERB Monolingual track. ``GroupWS'' indicates Grouped Weighted Sums and ``Hierarchical Conv.'' stands for Hierarchical Convolution over layers. The metric is WER/CER.}
  \label{tab:preliminary}
  \centering
  \begin{tabular}{ l c c c }
    \toprule
    Interface & Params & Mono-10min & Mono-1hr \\
    \midrule
    Weighted Sum & 13 & 42.85           & 35.15 \\
    \midrule
    GroupWS(\#Groups=2) & 1.2M & 41.84     & 34.47 \\
    GroupWS(\#Groups=3) & 1.8M & 43.08     & 33.96 \\
    GroupWS(\#Groups=4) & 2.4M & 42.52     & 33.99 \\
    Concat $+$ Proj & 7.7M & 43.20      & 34.26 \\
    PCA $+$ Concat & 0 & 45.16          & 36.58 \\
    Hierarchical Conv. & 4.4M & \textbf{41.51}   & \textbf{33.88} \\
    CLS Pooling & 5.5M & 48.36          & 33.92 \\

    \bottomrule
  \end{tabular}
  \vspace{-15pt}
\end{table}

\begin{table*}[th]
  \caption{Performance comparison of proposed interfaces using different upstream models, evaluated on both ML-SUPERB and SUPERB. The metric for each task: 
  ``Mono-1hr'' and ``Multi-1hr'': WER/CER,
  ``LID-1hr'',``ER'',``IC'', ``SV'': accuracy,
  ``PR'': PER.
  }
  \label{tab:full_results}
  \centering
  \begin{tabular}{ l l c c c c c c c }
    \toprule
    \multirow{ 2}{*}{Upstream}  & \multirow{ 2}{*}{Interface} & \multicolumn{3}{c}{ML-SUPERB} & \multicolumn{4}{c}{SUPERB}   \\
    &  & Mono-1hr~($\downarrow$) & Multi-1hr~($\downarrow$)  & LID-1hr~($\uparrow$)
    & ER~($\uparrow$) & IC~($\uparrow$) & SV~($\downarrow$) & PR~($\downarrow$) \\
    \midrule
    \multirowcell{ 3}{HuBERT\\Base} 
    &  Weighted Sum  &	35.1 & 24.4 & \textbf{84.6} & 64.92 & 98.34 & \textbf{5.11} & 5.41 \\
    &  Hierarchical Conv.  &	\textbf{33.9} & \textbf{24.0} & 81.7 & \textbf{68.49} & \textbf{99.45} & 5.62 & 3.07\\
    &  CLS Pooling &	\textbf{33.9} & 24.3 & 1.2 & 62.73 & \textbf{99.45} & 49.91 & \textbf{2.93} \\
    \midrule

    \multirowcell{ 3}{HuBERT\\Large} 
    &  Weighted Sum         &	32.3 & 22.3 & 64.4 & 67.62 & 98.76 & \textbf{5.98} & 3.53 \\
    &  Hierarchical Conv.    &	\textbf{30.0} & \textbf{21.4} & \textbf{89.2} & \textbf{72.44} & \textbf{99.53} & 6.03 & 1.76 \\
    &  CLS Pooling           &	31.1 & 21.5 & 87.0 & 68.72 & \textbf{99.53} & 6.34 & \textbf{1.64} \\
    \midrule

    \multirowcell{ 3}{WavLM\\Base} 
    &  Weighted Sum  &	34.2 & 24.3 & \textbf{84.1} & 65.94 & 98.63 & \textbf{4.69} & 4.84 \\
    &  Hierarchical Conv.  &	\textbf{32.4} & \textbf{23.6} & 67.9 & \textbf{68.57} & \textbf{99.53} & 5.48 & 3.06\\
    &  CLS Pooling  &	33.8 & 23.8 & 13.2 & 63.60 & 99.50 & 6.03 & \textbf{2.92}\\
    \midrule

    \multirowcell{ 3}{WavLM\\Large} 
    &  Weighted Sum  &	30.1 & 20.8 & 71.7 & 70.62 & 99.31 & \textbf{3.77} & 3.06\\
    &  Hierarchical Conv.  &	\textbf{28.0} & 19.4 & \textbf{90.6} & \textbf{74.95} & \textbf{99.71} & 5.20 & \textbf{1.72}\\
    &  CLS Pooling &	29.7 & \textbf{18.8} & 88.5 & 70.45 & 99.63 & 4.27 & 1.73 \\
    \midrule

    \multirow{ 3}{*}{XLSR-53} 
    &  Weighted Sum  &	35.1 & 20.2 & 76.6 & 66.34 & 95.62 & 6.45 & 4.50\\
    &  Hierarchical Conv.  &	\textbf{30.6} & \textbf{19.8} & \textbf{80.4} & \textbf{72.01} & \textbf{99.55} & \textbf{5.63} & \textbf{2.69}\\
    &  CLS Pooling &	31.0 & 80.7 & 4.5 & 65.10 & 99.47 & 12.95 & 5.27\\   

    \bottomrule
  \end{tabular}
  \vspace{-10pt}
\end{table*}

\subsection{Experimental Setup}

To get a holistic performance comparison of different interface designs, we choose 5 SSL models and 2 speech SSL benchmarks for evaluation.
Specifically, we pick HubERT Base and Large~\cite{hsu2021hubert}, WavLM Base and Large~\cite{Chen2021WavLMLS}, and XLSR-53~\cite{Conneau2020xlsr} for our upstream models.
For downstream tasks, we test all proposed interfaces on ML-SUPERB~\cite{shi23g_mlsuperb} because it is currently less saturated than many SUPERB tasks.
ML-SUPERB contains both 10min and 1hr training sets for each language.
In the ML-SUPERB Monolingual track, the overall reported score is the average CER across 13 languages (the model for each language is trained individually).
In the multilingual track, there are two tasks: \textbf{ASR} and Language Identification~(\textbf{LID}).
In these tasks, a model must be trained on 143 languages simultaneously.

In addition to multilingual ASR tasks, we also test the generalization of our interface designs on other common speech processing tasks.
Hence, we select 4 tasks from  SUPERB: Intent Classification~(\textbf{IC}), Phoneme Recognition~(\textbf{PR}), Emotion Recognition~(\textbf{PR}), and Speaker Verification~(\textbf{SV}).

To reduce the computational cost of running a large number of potential experiments, we first test all proposed interfaces on ML-SUPERB Monolingual Track with HuBERT Base.
Then we pick the top 2 most promising interface designs and run them on all 5 upstream models and both ML-SUPERB and SUPERB.
For both SUPERB and ML-SUPERB, we follow the default training configurations as far as possible.
However, for the Multilingual track on ML-SUPERB, we find that the default downstream model
is not optimal, and performance can be easily increased by simply adding more layers.
In order to disentangle the effect of a larger downstream model from the interface itself, we do a hyperparameter search on the best number of layers in the downstream model so that it can achieve the lowest CER in terms of the same frozen upstream model.
From our empirical experiment, we found that 65 is the optimal layer number for Multilingual ASR~(about 56M params).

\subsection{Interface Comparison on ML-SUPERB Monolingual ASR with HuBERT Base}

The results for our pilot experiment using HuBERT Base on the ML-SUPERB Monolingual task are shown in Table~\ref{tab:preliminary}. Firstly, we notice that there is a significant performance difference between all proposed interfaces relative to the baseline weighted sum method. While several of the proposed interfaces improve over the baseline weighted sum, we see that the Hierarchical Convolution achieves the overall best interface performance on both the 10min and 1hr datasets.
The CLS Pooling interface ranks second on the 1hr task, but performs poorly on 10min, indicating that this method likely requires more finetuning data to reach its potential.
In light of their promising results on the 1hr ML-SUPERB monolingual task, we choose Hierarchical Convolution and CLS Pooling for further experimentation.

\subsection{Interface Comparison on ML-SUPERB and SUPERB}
The results for the full suite of tasks across ML-SUPERB and SUPERB using all upstream models are shown in Table~\ref{tab:full_results}. Again, we observe that the Hierarchical Convolution interface tends to consistently outperform the baseline weighted sum as well as the CLS Pooling interface.
The performance differences are often substantial: for example, simply replacing the weighted sum interface with the Hierarchical Convolution interface reduces HuBERT Base's PER on the SUPERB phone recognition task from 5.41 to 2.93, which is lower than the 3.53 PER achieved by HuBERT Large with the weighted sum interface.
We even find that the CLS Pooling interface with the HuBERT Large model establishes a new SotA on the SUPERB leaderboard for phone recognition at 1.64 PER, beating the WavLM model's 3.06 PER when using the weighted sum.

Generally, the Hierarchical Convolution interface
outperform the baseline weighted sum on all tasks except for Speaker Verification with HuBERT and WavLM models, although XLSR-53 with Hierarchical Convolution outperforms the weighted sum.
This might imply that speaker-related information is encoded similarly across different layers in HuBERT and WavLM, causing our proposed interfaces to overfit.

Interestingly, in both benchmarks, we also observe a larger improvement over the basic weighted sum in Large models compared to Base models for HuBERT and WavLM when Hierarchical Convolution is applied, which supports our hypothesis that the basic weighted sum averages out more information as the number of layers increases.
In particular, on the LID task, we see that with Base upstream models the Hierarchical Convolution interface underperforms the basic weighted sum, but when scaling to the Large models (HuBERT Large, WavLM Large, and XLSR-53) the Hierarchical Convolution instead outperforms the basic weighted sum.

\subsection{Interface vs. Larger Downstream Models}

\begin{table}[th]
\scriptsize
\vspace{-5pt}
  \caption{Ablation experiments on trainable parameters. ``WS w/ Large DS'' indicates Weighted Sum with Large Downstream model.}
  \label{tab:ablation}
  \centering
  \begin{tabular}{ l l c c }
    \toprule
    \multirow{ 2}{*}{Upstream}  & \multirow{ 2}{*}{Interface} & \multicolumn{2}{c}{ML-SUPERB~(WER/CER)}    \\
    &  & Mono-1hr~($\downarrow$) & Multi-1hr~($\downarrow$) \\
    \midrule
    \multirowcell{ 2}{HuBERT\\Base} 
    &  Hierarchical Conv. &     \textbf{33.9} & \textbf{24.0} \\
    &  WS w/ Large DS &            35.2 & 24.8  \\
    \midrule

    \multirowcell{ 2}{HuBERT\\Large} 
    &  Hierarchical Conv.   &	\textbf{30.0} & \textbf{21.4}  \\
    &  WS w/ Large DS          &	32.9 & 22.1  \\
    \midrule

    \multirowcell{ 2}{WavLM\\Base} 
    &  Hierarchical Conv. &     \textbf{32.4} & \textbf{23.6} \\
    &  WS w/ Large DS &            34.4 & 24.3 \\
    \midrule

    \multirowcell{ 2}{WavLM\\Large} 
    &  Hierarchical Conv. &     \textbf{28.0} & \textbf{19.4} \\
    &  WS w/ Large DS &            30.5 & 20.8  \\
    \midrule

    \multirow{ 2}{*}{XLSR-53} 
    &  Hierarchical Conv. &     \textbf{30.6} & \textbf{19.8} \\
    &  WS w/ Large DS &            35.4 & 20.0 \\   

    \bottomrule
  \end{tabular}
  \vspace{-5pt}
\end{table}

Although there are substantial improvements using the Hierarchical Convolution interface, it is not immediately clear whether the performance gain comes from the interface design or simply the additional trainable parameters.
To this end, we increase the size of the downstream model for Monolingual and Multilingual ASR on ML-SUPERB by an amount roughly the same as the number of parameters in the Hierarchical Convolution interface (denoted as ``WS+Large DS'').

As shown in Table~\ref{tab:ablation}, under all settings, weighted sum with a large downstream model does not lead to better performance than the Hierarchical Convolution interface combined with a smaller downstream model.
This implies the role of the interface is not simply to add more learnable parameters to the downstream model, but to actually better aggregate information across the layers of the upstream model.
Our findings not only resonate with~\cite{zaiem23b_interspeech} that larger downstream models increase the performance but we also show that a better interface design can get further improvements.

\subsection{Performance under Fine-tuning setting}
We motivated the introduction of the interface module under the assumption that the upstream model would typically remain frozen during the entire downstream task fine-tuning.
However, we wish to empirically verify whether different interfaces can still offer improvements with end-to-end fine-tuning.
Practically, we follow the same pipeline as in our previous experiments except the upstream model is trainable.
We evaluate on ML-SUPERB Monolingual track 1h using HuBERT Base.

\begin{table}[th]
\scriptsize
\vspace{-5pt}
  \caption{Comparison of Weighted Sum and Hierarchical under end-to-end fine-tuning. The metric is WER/CER.}
  \label{tab:finetune}
  \centering
  \begin{tabular}{ l c c }
    \toprule
     \multirow{ 2}{*}{Interface} & \multicolumn{2}{c}{ML-SUPERB Mono-1hr~($\downarrow$)}    \\
      & Freeze Upstream & Fine-tuned \\
    \midrule
    
    Weighted Sum &     35.1 & 31.5 \\
     
     Hierarchical Conv. &  \textbf{33.9} & \textbf{31.1} \\
      WS w/ Large DS &       35.2 & 31.6 \\

    \bottomrule
  \end{tabular}
  \vspace{-5pt}
\end{table}

As shown in Table~\ref{tab:finetune}, end-to-end fine-tuning shortens the gap between different interface designs.
However, the Hierarchical Convolution interface still outperforms the weighted sum baseline.
These results indicate that the interface design is beneficial regardless of frozen or trainable upstream model.

\section{Conclusion}
\label{sec:conclusion}
In this work, we introduced the interface module into the conceptual framework for utilizing self-supervised speech models.
We also proposed several new interface designs.
From our experiments on multiple benchmarks and upstream models, we showed that the Hierarchical Convolution interface tends to outperform our other proposed designs, as well as the weighted sum method that is currently standard practice in the field.
Finally, our ablation experiments substantiate that the role of the interface in optimally aggregating information across layers of the upstream model is more important 
than the amount of trainable parameters during funetuning.


\section{Acknowledgements}
This work is supported in part by the National Science Foundation under Grant No. 2238605.


\bibliographystyle{IEEEtran}
\bibliography{mybib}

\end{document}